\documentclass[prb,twocolumn,superscriptaddress,showpacs]{revtex4}

\usepackage{graphicx}  
\usepackage{dcolumn}   

\begin{document}

\def\be{\begin{equation}} 
\def\ee{\end{equation}}
\def\ba{\begin{eqnarray}} 
\def\ea{\end{eqnarray}} 
\def\J{J}
\def\tJ{\tilde {J}}
\def\L{L}

\title{DMRG Study of Critical Behavior of the Spin-1/2 Alternating
Heisenberg Chain}

\author{T. Papenbrock} 
\affiliation{Physics Division, Oak Ridge National Laboratory, Oak
Ridge, TN 37831-6373}

\author{T. Barnes}
\affiliation{Physics Division, Oak Ridge National Laboratory, Oak
Ridge, TN 37831-6373}
\affiliation{Department of Physics and Astronomy, University of Tennessee, 
Knoxville TN 37996-1201}

\author{D. J. Dean}
\affiliation{Physics Division, Oak Ridge National Laboratory, Oak
Ridge, TN 37831-6373}

\author{M. V. Stoitsov}
\affiliation{Physics Division, Oak Ridge National Laboratory, Oak
Ridge, TN 37831-6373}
\affiliation{Department of Physics and Astronomy, University of Tennessee, 
Knoxville TN 37996-1201}
\affiliation{Institute for Nuclear Research and Nuclear Energy, Bulgarian Academy of Sciences, Sofia-1784, Bulgaria}
\affiliation{Joint Institute for Heavy Ion Research, Oak Ridge, TN 37831}

\author{M. R. Strayer} 
\affiliation{Physics Division, Oak Ridge National Laboratory, Oak
Ridge, TN 37831-6373}

\date{\today}

\begin{abstract}
We investigate the critical behavior of the $S=1/2$ alternating
Heisenberg chain using the density matrix renormalization group
(DMRG). The ground-state energy per spin 
${\tilde e}_0$ 
and
singlet-triplet energy gap 
${\tilde \Delta}$ 
are determined for a range of
alternations $\delta$.  Our results for the approach of
${\tilde e}_0$ 
to the uniform chain limit are well described
by a power law with exponent $p\approx 1.45$. The singlet-triplet gap
is also well described by a power law, with a critical exponent of
$p\approx 0.73$, half of the ${\tilde e}_0$ exponent. The 
renormalization group predictions of power laws with logarithmic
corrections can also accurately describe our data
provided that a surprisingly large scale parameter 
$\delta_0$ is present in the logarithm. 
\end{abstract}

\maketitle

\section{Introduction}

The approach of quantum spin systems to criticality is an interesting
and rather underexplored problem in computational physics.  In this
paper we
consider the critical behavior of one of the simplest quasi-1D quantum
spin systems, the alternating Heisenberg chain with spin-1/2, 
as it approaches the
uniform chain limit.
This model is of special interest for the study of
critical behavior, since much is known analytically 
about the gapless
uniform-chain limit through the Bethe Ansatz.
This system is also important for studies of the
magnetic spin-Peierls effect, and gives a reasonably accurate
description of magnetic interactions in many dimerized quasi-1D
antiferromagnets.

The alternating Heisenberg antiferromagnet is a simple generalization
of the uniform Heisenberg chain, with 
the nearest-neighbor exchange constant $J$ replaced
by two alternating values. 
The spin-1/2 system is defined by the
Hamiltonian
\begin{equation}
H = \tJ\sum_{i=1}^{\L/2} \left(
(1+\delta) \, \vec S_{2i-1}\cdot \vec S_{2i} 
+
(1-\delta) \, \vec S_{2i}\cdot \vec S_{2i+1}
\right) 
\end{equation}
and has a spin-singlet ground state with energy $E_0$ and a nonzero
singlet-triplet gap $E_1-E_0$ for any alternation\cite{Def} 
$0 < \delta < 1 $.  

For our study we introduce a finite-lattice,
scaled ground-state energy per spin,
\begin{equation}
\label{e0intro}
{\tilde e}_0(\L,\delta) =  E_0(\L,\delta)\; / \L \tJ \,
\end{equation}
and a corresponding singlet-triplet gap
\begin{equation}
\label{gapintro}
\tilde \Delta(\L,\delta) =  
\Big( E_1(\L,\delta) - E_0(\L,\delta) \Big)\; /\, \tJ   \ .
\end{equation}
The tilde indicates that these energies are scaled by 
$ \tJ $ rather than $\J\equiv (1+\delta)\tJ$, which would divide our 
results by a factor of $(1+\delta)$ ({\it albeit} giving the same 
leading critical behavior). In addition, 
energies quoted with a single real argument
and a tilde
({\it e.g.} $\tilde e_0(\delta)$) are bulk limits
of the scaled alternating chain, 
those with no tilde and a single integer argument 
({\it e.g.} $e_0(L)$) are finite-$L$ uniform chain results,
and those with no arguments
({\it e.g.} $e_0$) are bulk limits
of the uniform chain.

The uniform spin-1/2 Heisenberg antiferromagnetic chain, which we
recover at $\delta=0$, is the best understood 1D critical
quantum spin system. It has a ground-state energy per spin of
$e_0 = 1/4 - \ln 2$ and a band of gapless
spin-triplet excitations with dispersion relation $\omega(k)/\J =
\pi/2$ $|\sin(k)|$. The approach of noncritical models to this
limiting case is not well established, and has been the subject of
surprisingly few theoretical and numerical studies.

Early analytical studies of
the effect of a small alternation $\delta$ on the uniform chain 
were reported by Cross and Fisher \cite{Cro79}
and Black and Emery \cite{Bla81}. Cross and
Fisher used a Jordan-Wigner transformation to map the original spin
problem onto a pseudo-fermion Hamiltonian, and approximated the latter
by the exactly solvable Luttinger-Tomanaga model. This approach, which
unfortunately involves uncontrolled approximations, yields critical
exponents (defined by $f(\delta) \propto \delta^p$) 
of $p=4/3$ for the ground-state energy and $p=2/3$ for the
singlet-triplet gap. 

Black and Emery \cite{Bla81} related the critical
behavior of the alternating 
Heisenberg chain to the 4-state Potts model, and found
logarithmic corrections to these power laws,
\be
\label{theo_e0}
e_0-{\tilde e}_0(\delta)
\propto \delta^{4/3}/|\log{\delta}\, |
\ee 
for the ground-state energy per spin, and
\be
\label{theo_gap}
\tilde \Delta(\delta)\propto \delta^{2/3}/|\log{\delta}\, |^{1/2} 
\ee 
for the gap.
Note that at this order the gap scales as the
square root of the ground-state energy defect. More recent theoretical work
by Affleck {\it et al.} \cite{Affleck89} 
has shown that 
Eqs.~(\ref{theo_e0},\ref{theo_gap}) are 
leading-order predictions of the 
renormalization group. The overall constants in these
results and implicit in the logarithms are non-universal,
and have not yet
been determined analytically for this model. 

Several previous numerical studies have 
investigated the critical behavior of
spin-1/2 alternating Heisenberg chains using bulk-limit extrapolations
of exact diagonalization results on systems up to about $L=30$ 
in extent (see Barnes {\it et al.} \cite{Bar99} 
and Yu and Haas \cite{Haas} and references cited therein).  
We shall see that 
important systematic errors 
can arise from extrapolations using
these relatively small systems, for example
in estimates of critical exponents.

Studies of much larger 
spin-1/2 alternating Heisenberg chains 
have also been 
published using the DMRG algorithm, 
although there has been little 
systematic study of the critical behavior of the simple alternating chain
model of Eq.(1) using DMRG.  The single published reference on this
topic is the work of Uhrig {\it et al.} \cite{Uhrig99}, who estimate
a gap critical exponent of $0.65$. Their numerical energies however
deviate systematically from this power law at small alternation
$\delta$ (see Fig.3 of Uhrig {\it et al.} \cite{Uhrig99}).  
In related work, Chitra
{\it et al.} \cite{Chitra} used the DMRG method 
to study the effects of
dimerization and frustration on a generalized alternating chain model
with next-nearest-neighbor couplings, and Lou {\it et al.} \cite{Lou}
studied the gap induced by a staggered magnetic field.

In this work we present a systematic DMRG study of the critical
behavior of the original alternating spin-1/2 Heisenberg chain of
Eq.(1), and compare the theoretical predictions Eqs.(4,5) to
numerical results for the ground-state energy per spin and
singlet-triplet gap of this model, using systems up to $L=192$ in
extent.

\section{Numerical method}

For our numerical study we employed the
density matrix renormalization group (DMRG) algorithm \cite{White},
which
is a very effective method for studying critical behavior in
quasi-1D quantum spin systems. 
Numerical determination of the critical behavior of the alternating
chain, or any similar quantum spin system, is a daunting computational
task. One must accurately determine energy eigenvalues on quite large
systems, since characteristic lengths typically diverge at critical
points. An extrapolation through a series of fixed-$L$
results is then required on sufficiently large lattices to insure that
one is in an asymptotic regime in which finite size effects can be
accurately parametrized and eliminated. The computer memory
requirements for diagonalizing these large systems are such that
the detailed critical behavior of relatively few 
quantum spin systems has been explored 
numerically.

The DMRG algorithm 
has previously been applied to various interacting fermion
systems, including one-dimensional spin chains \cite{White}, 
lattice models \cite{Pes99}, quasilinear molecules \cite{Ziosi} and
nuclei \cite{Duk01,Dim02}. The essential concept in DMRG is to
``grow'' a small, finite system into a larger one by the iterative
incorporation of new lattice sites. At each such iteration one 
retains only
the $m$ most relevant basis states 
for spanning the targeted energy eigenstate.
(These basis states are chosen according to a 
density-matrix weight.) 
This selective sampling of
Hilbert space yields accurate energy eigenvalues on systems
which are well beyond the limits of exact
diagonalization.  For details of the DMRG method we refer the reader
to the original papers of White \cite{White} and to a series of
lectures recently compiled by Peschel {\it et al.}  \cite{Pes99}

Our numerical implementation proceeds as follows.  We divide the spin
chain into blocks $A-a-B-b$, where $A$ and $B$ denote the ``system''
and ``environment'' blocks, and $a$ and $b$ are elementary blocks to
be added to $A$ and $B$ respectively. Blocks $b$ and $A$ are
linked by periodic boundary conditions. We take spin dimers (two
lattice sites of spin $s=1/2$ each) as our elementary blocks, and use
the infinite algorithm to grow the spin chain, while targeting the
lowest-lying spin-0 or spin-1 state. In the case of the spin-0 ground
state of the alternating Heisenberg chain, we found that subsequent
sweeps with the finite algorithm do not lead to much improvement 
in the results of the infinite algorithm. In contrast, 
for the spin-1 state we found that sweeps with the finite
algorithm were important for convergence. 
Our DMRG implementation uses large, sparse matrices, so the sparse
matrix package {\sc Arpack} \cite{Arpack} was employed. 
The time consuming
matrix-vector multiplications and the Arnoldi/Lanczos algorithm {\sc
Arpack} were parallelized.
 
The ground-state
energy per spin ${\tilde e}_0(\L,\delta)$ of a spin chain of 
length $L$
and alternation $\delta$ 
is of special interest for our study of critical behavior 
(see Eq.~(\ref{e0intro})). DMRG yields
approximate values ${\tilde e}_0(m;\L,\delta)$ that
typically converge exponentially fast from above as one increases the number 
$m$ of states retained \cite{White,Pes99}. We also observed this
behavior in the present study.  We computed ${\tilde e}_0(m;\L,\delta)$
numerically using DMRG on chains of length $L=28$, 48, 96, 144 and 192,
and alternation $\delta= 2^{-3},
2^{-4},2^{-5},\ldots,2^{-10}$. 
At each $L$ and $\delta$ we increased  
$m$ in steps of 10 starting at $m=30$ until a fit of
the form ${\tilde e}_0(m;{\L},\delta)={\tilde e}_0({\L},\delta)
+c_1\exp{(-c_2 m)}$ gave sufficiently 
stable coefficients for our desired accuracy; 
this $m$ extrapolation yielded our DMRG
energy estimate ${\tilde e}_0(\L,\delta)$. We found that the maximum
$m$ needed for convergence to a given accuracy increases with
increasing chain length $L$ and decreasing alternation $\delta$.  For
our extreme case $L=192$ and $\delta=2^{-10}$, adequate convergence was
not achieved until $m=150$, and we retained the maximum of $m=170$
states in this case. This resulted in a sparse matrix problem of
dimension $\approx 9\times 10^4$ at each DMRG iteration.
Finally, we also confirmed recovery of 
exact $L=28$ alternating chain results on allowing our DMRG code 
to iterate to the full
Hilbert space.

\section{Results}

Table~\ref{tab1} gives our DMRG results for the ground-state
energy per spin for different alternations $\delta$ and chain lengths
$L$. For the rather large alternations $\delta=2^{-3}$ and $2^{-4}$ we
confirmed convergence to accurately known 
energies with increasing basis state number $m$,
as well as convergence with increasing chain
length $L$. 
Finite size effects were more
pronounced at smaller $\delta$, 
as expected since the system is closer to criticality.  
We
estimate that the energy errors in Table~\ref{tab1} 
are a few units in the last digit, 
based on the difference between the
maximum-$m$ DMRG result and the exponential 
$m\to \infty$ extrapolation which we quote.

\begin{figure}
\includegraphics[width=0.47\textwidth]{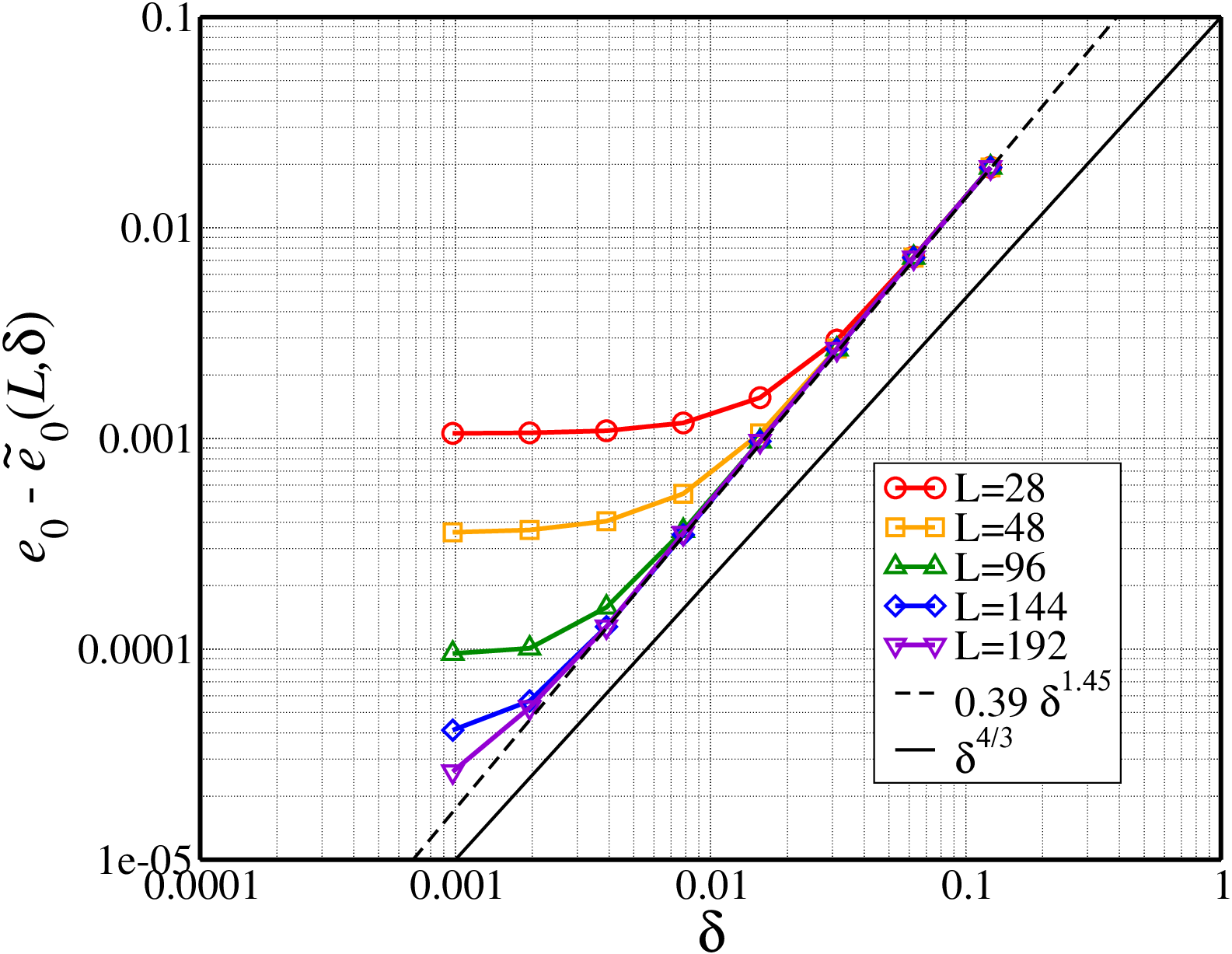}
\protect\caption{\label{fig1}Approach of 
${\tilde e}_0(\L,\delta)$ to the
uniform-chain limit $e_0 = 1/4 - \ln(2)$ with decreasing alternation
$\delta$ and increasing chain length $L$.
The dashed line shows a fit to a power law,
$c_0 \delta^p$, which gives $c_0 = 0.39 $ and an exponent of
1.45. The full
line shows the renormalization-group 
exponent 4/3 (displaced for presentation, 
without logarithmic corrections).}
\end{figure}

Fig.~\ref{fig1} shows the approach of ${\tilde e}_0(\L,\delta)$ to the
limit $e_0 = 1/4-\ln{2}$ as a function of alternation $\delta$ for
spin chains of increasing length, $L=28$, 48, 96, 144 and 192.  The
asymptotic large-$L$ envelope evident in this figure for $\delta \agt
2^{-8}$ clearly shows the bulk limit.  The dashed line is a power-law
fit $c_0\delta^p$ to this envelope, which gives $c_0=0.39$ and
exponent $p=1.45$.  The renormalization-group prediction $p=4/3$ from
Eq.~(\ref{theo_e0}) (without the logarithmic factor) is also shown,
displaced for clarity of presentation. Clearly this gives an inferior
description of the DMRG data over the range considered here.

\begin{figure}
\includegraphics[width=0.47\textwidth]{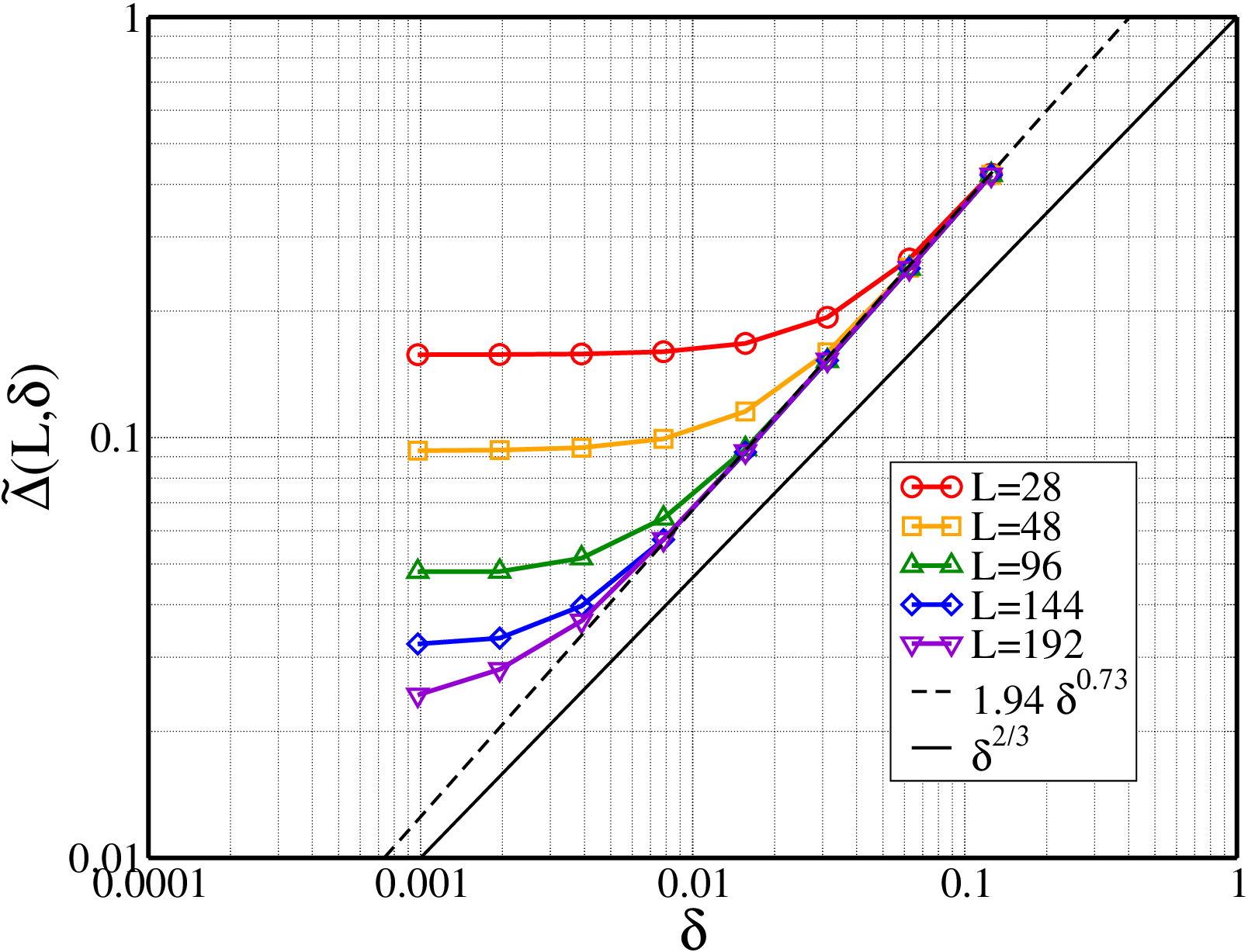}
\protect\caption{\label{fig2}Critical behavior of 
the singlet-triplet gap
$\tilde \Delta(\L,\delta)$ as a function of alternation $\delta$
for spin chains of length $L$. The dashed line shows
a power law fit, $c_0\delta^p$, which gives $c_0=1.94$ and exponent $p=0.73$. 
The full line shows the renormalization-group
exponent $p=2/3$ (displayed as in Fig.~\ref{fig1}).
}
\end{figure}

Fig.~\ref{fig2} shows the singlet-triplet gap ${\tilde \Delta}(L,\delta)$ 
as a
function of alternation $\delta$ for spin chains of
length $L=28$, 48, 96, 144 and 192. 
The envelope of these curves
is the bulk limit 
${\tilde \Delta}(\delta)$,  
which is 
evident for $\delta \agt 2^{-7}$. 
A power-law fit
to ${\tilde \Delta}(\delta)$ as for ${\tilde e}_0(\delta)$ gives
$c_0 = 1.94$ and $p=0.73$, 
shown as a dashed line in Fig.~\ref{fig2}. 
The renormalization-group prediction $p=2/3$ from
Eq.(\ref{theo_gap})
is also shown, displaced for presentation. 
Evidently the renormalization-group exponent (without the logarithmic
term)
again gives a less accurate description of our DMRG data. Note that
the two exponents obtained in our fits are related by a
factor of two, which implies that the 
${\tilde e}_0(\delta)$ defect scales as
${\tilde \Delta}(\delta)^2$. 
This relation also follows from the leading-order
renormalization group, Eqs.~(\ref{theo_e0},\ref{theo_gap}).

It is especially interesting to determine whether there is numerical 
evidence for  
logarithmic corrections to pure power-law behavior, as predicted
by the renormalization group in
Eqs.(\ref{theo_e0},\ref{theo_gap}). We first consider the bulk-limit 
ground-state energy per spin, predicted 
to asymptotically approach the uniform chain
limit as 
\be
\label{trial}
e_0-{\tilde e}_0(\delta)
=\alpha{\delta^{4/3}\over \ln{\left(\delta/\delta_0\right)}}.
\ee
Here we have introduced an overall constant $\alpha$ and scale
parameter $\delta_0$, which we will estimate from our DMRG data.
On rearranging Eq.~(\ref{trial}) we obtain
the easily visualized form
\be
\label{plot}
{\delta^{4/3}\over {e}_0-{\tilde e}_0(\delta)}
=\alpha^{-1} \Big( \ln{\delta} - \ln{\delta_0}\Big).
\ee

\begin{figure}
\includegraphics[width=0.47\textwidth]{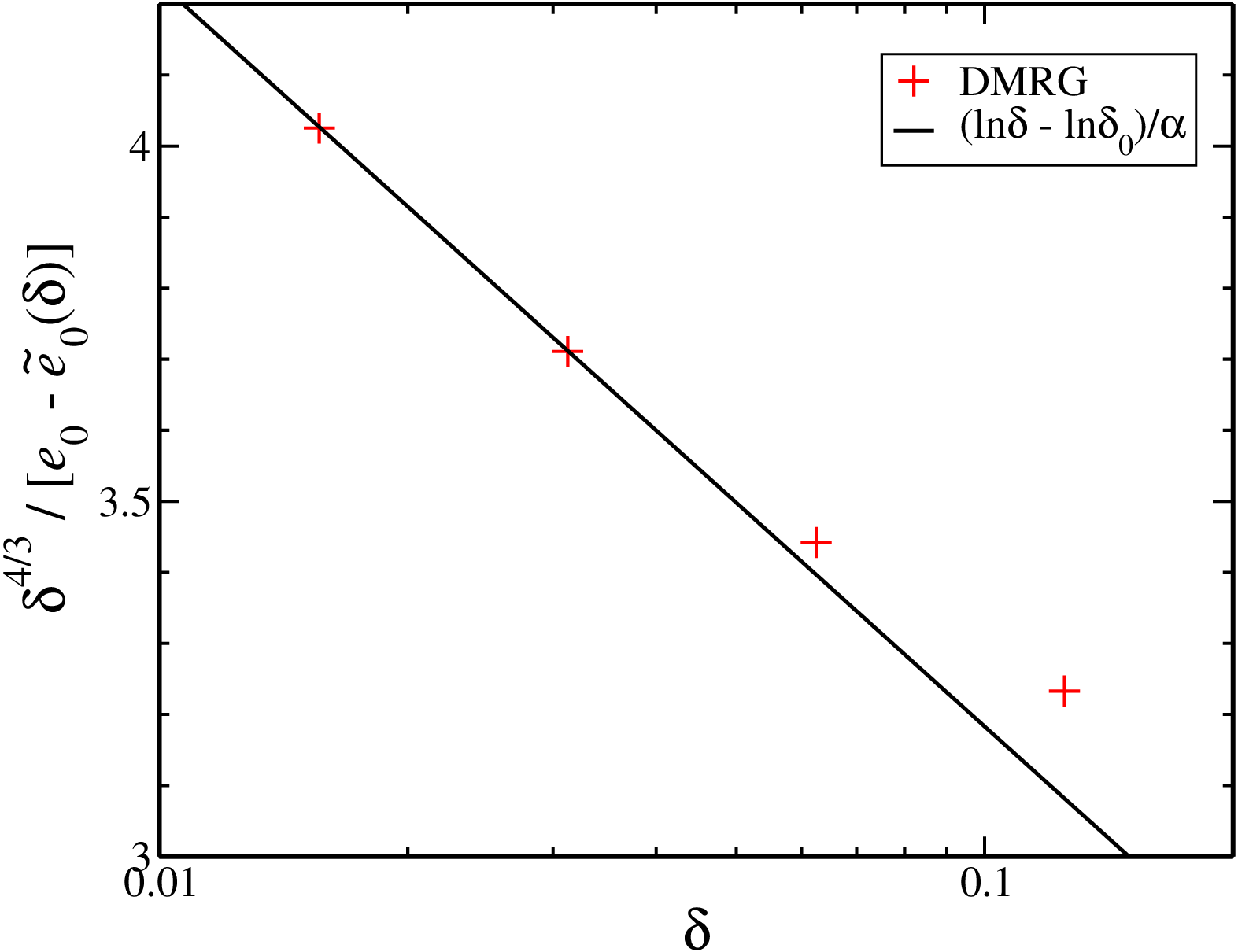}
\protect\caption{\label{fig3}Estimation of the constants
$\alpha$ and $\delta_0$ in Eq.~(\ref{trial}) from our DMRG data. 
}
\end{figure}

Fig.~\ref{fig3} shows our DMRG results for 
$\delta^{4/3}/({
e}_0-{e}_0(\delta))$ (the {\it l.h.s.}
of Eq.~(\ref{plot})) versus $\ln\delta$. The data
clearly disagree with the 
leading-order renormalization group prediction Eq.~(\ref{trial}) 
over this range
of $\delta$, since the points do not lie on a straight line. 
Assuming that 
the two smallest-$\delta$ points are close to asymptotic, 
we estimate
$\alpha\approx -2.2$ and $\delta_0\approx 110$.  
A similar fit to the singlet-triplet gap gives
\be
\left({\delta^{2/3}\over \tilde{\Delta}(\delta)}\right)^2
=\alpha_{gap}^{-1} \Big( \ln{\delta} - \ln{\delta_0}\Big)
\ee
with $\alpha_{gap}\approx -19.4$ and $\delta_0\approx 115$. 
With these constants the theoretical
renormalization-group results Eqs.~(\ref{theo_e0},\ref{theo_gap})
are barely distinguishable from our power-law fits  
in Figs.~\ref{fig1} and \ref{fig2}. There are no predictions of these constants 
in the literature to our knowledge. 

Eq.~(\ref{trial}) can be reexpressed as an effective power 
$p_{\it eff}(\delta) = 4/3+1/{\ln(\delta_0/\delta)}$. With 
$\delta_0 = 110$, the range $\delta = 10^{-3}\to 10^{-1}$ corresponds to
$p_{\it eff}=1.42\to 1.47$, which may explain our good numerical agreement
with a pure power of exponent $p=1.45$.

\begin{figure}
\includegraphics[width=0.47\textwidth]{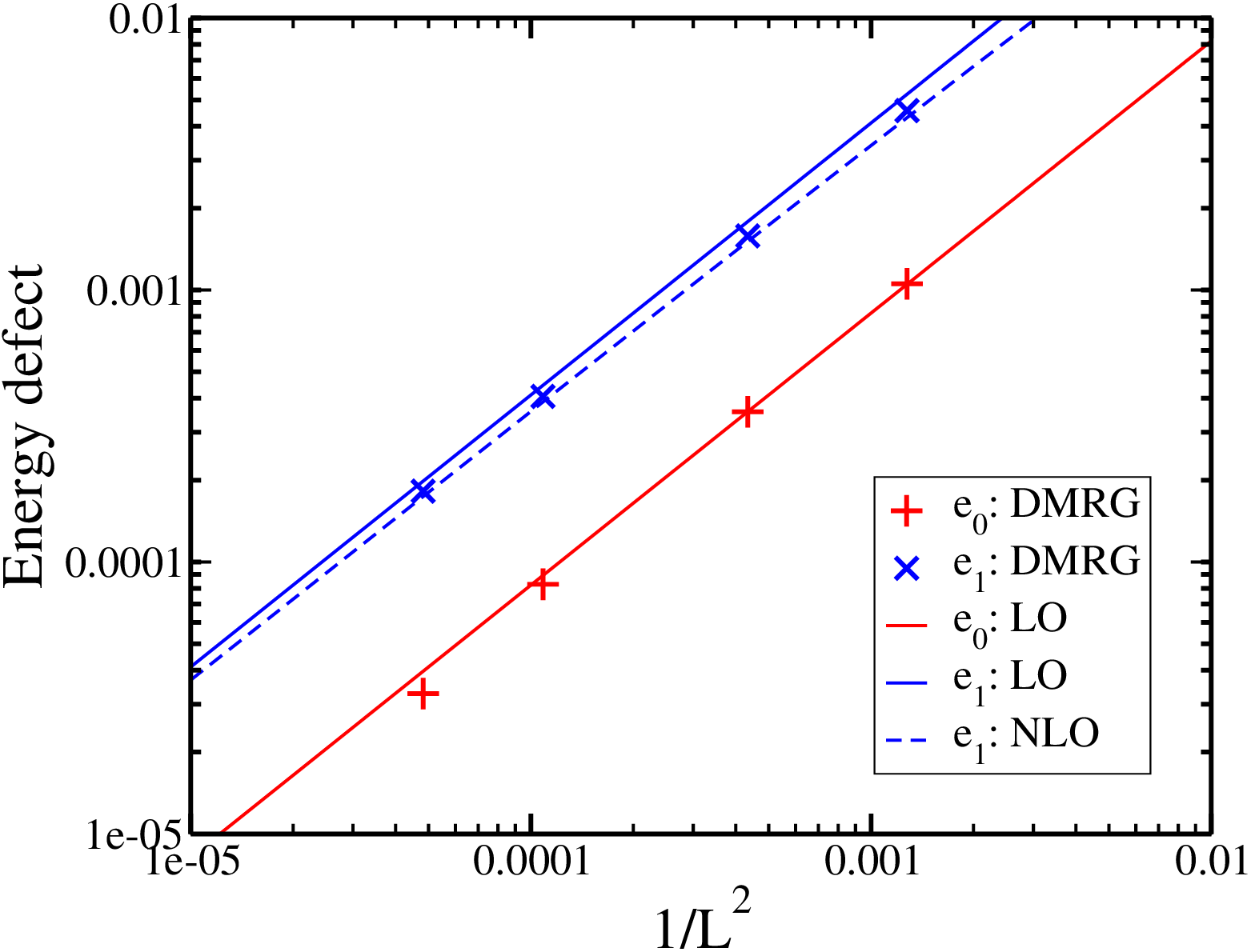}
\protect\caption{\label{fig4}Finite-size 
energy defects $|e_s(\L)-e_0|$ of the lowest spin-$s$ levels
on the uniform chain 
versus chain length $L$. The data points
are DMRG ($+$:spin-0, $\times$:spin-1) and the lines are 
Bethe-Ansatz results
(leading order (LO), solid; next-to-leading order
(NLO), dashed), from Eqs.(\ref{pred0},\ref{pred1}).
}
\end{figure}

We have also investigated finite size effects in the uniform Heisenberg chain 
($\delta=0$).
Woynarovich and Eckle \cite{WE87} and Affleck
{\it et al.}\cite{Affleck89} quote Bethe-Ansatz predictions for the 
leading finite size contributions 
to the ground-state energy per spin
and singlet-triplet gap,
\be
\label{pred0}
e_0-e_0(\L)={\pi^2\over 12\L^2} 
\Big(1 + O((\ln \L)^{-3})\Big)
\ee
and
\be
\label{pred1}
e_1(\L)-e_0
=
{5\pi^2\over 12\L^2}\Big(1-{3\over 5}\, (\ln \L )^{-1} +O((\ln \L)^{-2})\Big). 
\ee
Fig.~\ref{fig4} shows our DMRG data for these energy defects
on uniform
chains of length $L=28$, 48, 96 and 144, together with the Bethe-Ansatz
results.

The agreement between the leading-order Bethe-Ansatz predictions 
for finite size corrections to
the ground-state energy per spin (Eq.~(\ref{pred0})) and our
DMRG results is evidently very good on the shorter chains.
The discrepancy evident at $L=144$ may be due to convergence 
problems encountered by the DMRG algorithm 
when applied to this asymptotically gapless
system on large lattices.
The DMRG results for the
spin-1 energy defect depart significantly from the leading-order 
term in Eq.(\ref{pred1}), but are consistent with this
prediction when we include the $O(L^{-2}(\ln L)^{-1})$ correction.

Finally we consider the dependence of the ground-state energy 
per spin ${\tilde
e}_0(\L,\delta)$ on the chain length $L$ for fixed, nonzero alternation
$\delta$.  This is especially relevant to 
exact diagonalization studies, which  
extrapolate to the bulk limit from rather small systems of 
at most about
30 spins.  
To test the accuracy of finite size
extrapolations we used DMRG to compute the ground-state
energy per spin and singlet-triplet gap on chains of length
$L=28$, 32, 36, 40, 44 and 48, and fitted the results to
$f(\L)=a+b\exp{(-\L/c)}/\L$ and $g(\L)=a+b\exp{(-\L/c)}$
respectively. (These forms
were used to extrapolate exact diagonalization results
to the bulk limit by Barnes {\it et al.}\cite{Bar99}
and Yu and Haas.\cite{Haas})  
Both functions yield
reasonably good fits to our six data points in 
the range $L=28,\ldots, 48$.  
However, on 
comparing this extrapolation with our DMRG results on 
$L=96$, 144 and 192 lattices, we noted clear discrepancies
as the alternation $\delta$ decreases. 
This is likely due to a rapid increase in the characteristic length
(modeled by $c$ in the exponents of
$f(\L)$ and $g(\L)$),
which makes subleading terms in the asymptotic behavior more
important. Our results thus suggest caution in attempting
to establish critical behavior from studies of relatively
small systems.

\section{Summary}
We have employed the DMRG algorithm to determine 
the ground-state energy per spin and singlet-triplet gap
of the alternating Heisenberg chain, and studied the critical behavior
of this model in the limit of small alternation.
We find that the approach of the bulk-limit ground-state
energy per spin and singlet-triplet gap to the uniform chain limits 
are well described by power laws
in the alternation parameter $\delta$
over the range $0.008\alt\delta\alt 0.1$, and scale approximately as
$\delta^{1.45}$ and $\delta^{0.73}$ respectively.

The renormalization group predictions of
power laws times logarithmic corrections also 
appear consistent with our
results, provided that a surprisingly large scale factor $\delta_0$ 
is present in the logarithms.

\section*{Acknowledgments}
We acknowledge useful communications with I. Affleck, V. Barzykin,
G. S. Uhrig and S. R. White.  This research was sponsored in part by
the Laboratory Directed Research and Development Program of Oak Ridge
National Laboratory (ORNL), and computational resources were provided
by the Center for Computational Sciences at ORNL. ORNL is managed by
UT-Battelle, LLC for the U.S. Department of Energy under Contract
DE-AC05-00OR22725.

\begin{table}
\begin{ruledtabular}
\begin{tabular}{|d|d|d|d|}
\delta & \L & e_0(\L,\delta) & e_1(\L,\delta)\\\hline\hline
2^{-3}  &   28      &  -0.4110961       &  -0.3976678 \\     
2^{-3}  &   48      &  -0.4110928       &  -0.4032882\\
2^{-3}  &   96      &  -0.4110928       &  -0.407191 \\
2^{-3}  &   144     &  -0.4110928       &  -0.4084913 \\  
2^{-3}  &   192     &  -0.41109284      &  -0.4091416 \\\hline
2^{-4}  &   28      &  -0.4239186       &  -0.4149855\\      
2^{-4}  &   48      &  -0.4238627       &  -0.418895 \\
2^{-4}  &   96      &  -0.4238617       &  -0.4213883\\
2^{-4}  &   144     &  -0.4238619       &  -0.422211 \\   
2^{-4}  &   192     &  -0.4238618       &  -0.422623 \\  \hline  
2^{-5}  &   28      &  -0.4325593       &  -0.4258695 \\     
2^{-5}  &   48      &  -0.4323071       &  -0.429086 \\
2^{-5}  &   96      &  -0.4322899       &  -0.430754 \\
2^{-5}  &   144     &  -0.432290        &  -0.431263 \\   
2^{-5}  &   192     &  -0.4322906       &  -0.431518 \\ \hline   
2^{-6}  &   28      &  -0.4378673       &  -0.4319820\\      
2^{-6}  &   48      &  -0.437370        &  -0.435005 \\
2^{-6}  &   96      &  -0.437284        &  -0.436321 \\
2^{-6}  &   144     &  -0.437283        &  -0.436652 \\   
2^{-6}  &   192     &  -0.437285        &  -0.43681  \\ \hline    
2^{-7}  &   28      &  -0.4408870       &  -0.435219 \\      
2^{-7}  &   48      &  -0.440254        &  -0.438202 \\
2^{-7}  &   96      &  -0.440074        &  -0.439410 \\
2^{-7}  &   144     &  -0.440058        &  -0.439664 \\   
2^{-7}  &   192     &  -0.440064        &  -0.439768 \\  \hline   
2^{-8}  &   28      &  -0.4425056       &  -0.4368847 \\     
2^{-8}  &   48      &  -0.441826        &  -0.439863 \\
2^{-8}  &   96      &  -0.44158         &  -0.4410442\\
2^{-8}  &   144     &  -0.441550        &  -0.441275 \\   
2^{-8}  &   192     &  -0.441550        &  -0.441360 \\ \hline   
2^{-9}  &   28      &  -0.4433439       &  -0.4377293 \\     
2^{-9}  &   48      &  -0.442650        &  -0.440709 \\
2^{-9}  &   96      &  -0.442384        &  -0.441885 \\
2^{-9}  &   144     &  -0.442340        &  -0.442109 \\   
2^{-9}  &   192     &  -0.442336        &  -0.442190 \\  \hline  
2^{-10}  &  28      &  -0.4437703       &  -0.4381546\\      
2^{-10}  &  48      &  -0.443073        &  -0.441136\\
2^{-10}  &  96      &  -0.44281         &  -0.442311\\
2^{-10}  &  144     &  -0.442756        &  -0.442532\\
2^{-10}  &  192     &  -0.442741        &  -0.442614\\\hline
0       &   28      &  -0.444201        &  -0.4385820\\
0       &   48      &  -0.443504        &  -0.441566\\
0       &   96      &  -0.44323         &  -0.442741\\
0       &   144     &  -0.44318         &  -0.442965\\
\end{tabular}
\end{ruledtabular}
\caption\protect{\label{tab1}
Extrapolated DMRG results for the lowest 
energy per spin
in the spin-$s$ sector, 
$e_s(\L,\delta)\equiv {\tilde e}_s(\L,\delta)/(1+\delta)$, for various
alternations $\delta$ and chain lengths $L$.  The estimated error is
a few units in the final digit. 
}
\end{table}


\begin{thebibliography}{99}
\bibitem{Def}
A frequently used equivalent definition has alternating couplings
of $(\J,\J\alpha)  = 
( 
(1+\delta) \tJ,  
(1-\delta) \tJ 
)$.
\end{thebibliography}
\end{document}